\documentclass[%
 aip,
 amsmath,amssymb,
 reprint,%
]{revtex4-2}
\usepackage[english]{babel}
\usepackage[utf8]{inputenc}
\usepackage{xcolor}
\usepackage[colorinlistoftodos, color=green!40, prependcaption]{todonotes}
\usepackage{amsthm}
\usepackage{mathtools}
\usepackage{physics}
\usepackage{xcolor}
\usepackage{graphicx}
\usepackage[left=23mm,right=13mm,top=35mm,columnsep=15pt]{geometry} 
\usepackage{adjustbox}
\usepackage{placeins}
\usepackage[T1]{fontenc}
\usepackage{lipsum}
\usepackage{csquotes}

\setuptodonotes{inline,nolist}
\usepackage[pdftex, pdftitle={Article}, pdfauthor={Author}]{hyperref} 
\makeatletter
\def\@email#1#2{%
 \endgroup
 \patchcmd{\titleblock@produce}
  {\frontmatter@RRAPformat}
  {\frontmatter@RRAPformat{\produce@RRAP{*#1\href{mailto:#2}{#2}}}\frontmatter@RRAPformat}
  {}{}
}%
\makeatother

\begin{document}
\title{Gate controlled quantum dots in monolayer WSe\textsubscript{2}}

\author{Justin Boddison-Chouinard}
    \affiliation{Department of Physics, University of Ottawa, Ottawa, Ontario, K1N 9A7, Canada}
\author{Alex Bogan}
    \affiliation{Emerging Technologies Division, National Research Council of Canada, Ottawa, Ontario, K1A 0R6, Canada}
\author{Norman Fong}
    \affiliation{Emerging Technologies Division, National Research Council of Canada, Ottawa, Ontario, K1A 0R6, Canada}
\author{Kenji Watanabe}
    \affiliation{Research Center for Functional Materials, National Institute for Materials Science, 1-1 Namiki, Tsukuba 305-0044, Japan}
\author{Takashi Taniguchi}
    \affiliation{International Center for Materials Nanoarchitectonics, National Institute for Materials Science,  1-1 Namiki, Tsukuba 305-0044, Japan}
\author{Sergei Studenikin}
    \affiliation{Emerging Technologies Division, National Research Council of Canada, Ottawa, Ontario, K1A 0R6, Canada}
\author{Andrew Sachrajda}
    \affiliation{Emerging Technologies Division, National Research Council of Canada, Ottawa, Ontario, K1A 0R6, Canada}
\author{Marek Korkusinski}
    \affiliation{Emerging Technologies Division, National Research Council of Canada, Ottawa, Ontario, K1A 0R6, Canada}
\author{Abdulmenaf Altintas}
    \affiliation{Department of Physics, University of Ottawa, Ottawa, Ontario, K1N 9A7, Canada}
    \author{Maciej Bieniek}
     \affiliation{Department of Physics, University of Ottawa, Ottawa, Ontario, K1N 9A7, Canada}
    \affiliation{Department of Theoretical Physics, Wroclaw University of Science and Technology, Wroclaw, Poland}
    \author{Pawel Hawrylak}
    \affiliation{Department of Physics, University of Ottawa, Ottawa, Ontario, K1N 9A7, Canada}
\author{Adina Luican-Mayer$^{*}$}
    \affiliation{Department of Physics, University of Ottawa, Ottawa, Ontario, K1N 9A7, Canada}

\author{Louis Gaudreau$^{*}$}
    \affiliation{Emerging Technologies Division, National Research Council of Canada, Ottawa, Ontario, K1A 0R6, Canada}
\email[Correspondence email address: ]{luican-mayer@uottawa.ca, Louis.Gaudreau@nrc-cnrc.gc.ca }
\date{\today}

\begin{abstract}
Quantum confinement and manipulation of charge carriers are critical for achieving devices practical for quantum technologies. The interplay between electron spin and valley, as well as the possibility to address their quantum states electrically and optically, make two-dimensional (2D) transition metal dichalcogenides an emerging platform for the development of quantum devices. In this work, we fabricate devices based on heterostructures of layered 2D materials, in which we realize gate-controlled tungsten diselenide ($\mathrm{WSe_2}$) hole quantum dots.  We discuss the observed mesoscopic transport features related to the emergence of quantum dots in the $\mathrm{WSe_2}$ device channel, and we compare them to a theoretical model. 
\end{abstract}


\maketitle

One of the most promising solid-state platforms for implementing
qubits is based on semiconductor spin qubits utilizing
the spin degree of freedom of electrostatically confined
electrons in semiconductors \cite{brum_hawrylak_SLM1997,Loss1998}. This topic has been extensively studied in materials such as GaAs and SiGe, in which researchers have been able to create circuits from one to up to nine quantum dots \cite{Ciorga2000, Elzerman2003,Gaudreau2009,Zajac2016}, and have successfully demonstrated coherent quantum operations using the spin of the confined carriers\cite{Nichol2017, Zajac2018}. Despite significant progress, material specific challenges continue to hamper the development of complex quantum devices. For example, GaAs spin qubits suffer from short coherence times mainly due to hyperfine interaction with nuclei \cite{Hanson2007RMP}. SiGe devices have significantly higher qubit coherence times when using purified $^{28}$Si\cite{Veldhorst2014}, which has zero nuclear spin. However, the indirect band gap in Si prevents such devices from performing coherent photon to spin conversion, necessary for long distance quantum network communications \cite{Lodahl2017}.
\begin{figure*}
    \centering
    \includegraphics[width=\textwidth]{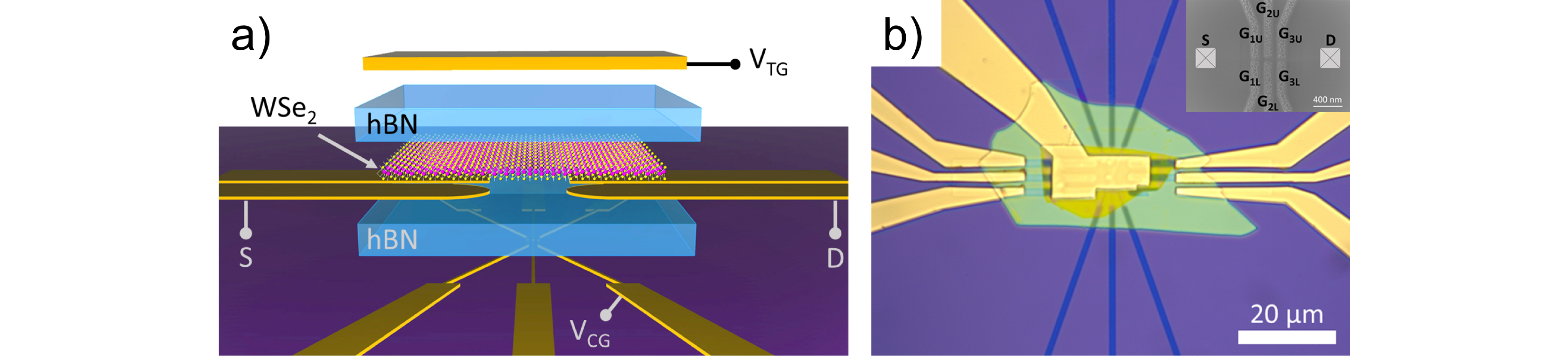}
    \caption{(a) Schematic of the monolayer $\mathrm{WSe_2}$ device structure. Cr (2 nm) / Pt (8 nm) contacts are used to contact monolayer $\mathrm{WSe_2}$ from the bottom. HBN encapsulates the $\mathrm{WSe_2}$ layer, while a top gate ($\mathrm{V_{TG}}$) and bottom control gates ($\mathrm{V_{CG}}$) are used to control the carrier density in the $\mathrm{WSe_2}$ layer. (b) Optical micrograph of a competed device. The inset represents a scanning electron micrograph depicting the layout of the control gates.}
    \label{fig:fig1}
\end{figure*}

With the development of advanced methods for fabricating 2D materials heterostructures, devices based on atomically thin semiconducting transition metal dichalcogenides (TMDs)\cite{Geim2013} are emerging as a promising platform in the field of quantum information science \cite{atature2018material, liu2019qis}. Semiconducting TMDs typically exhibit a direct band gap at the K-point in the Brillouin zone\cite{Mak2010}. Moreover, they have potential for realizing spin qubits with long coherence time due to spin-valley locking, caused by the strong spin-orbit interaction and the symmetries of their 2D lattice \cite{kormanyos2014spin}. These promising properties motivated recent efforts to understand charge transport through quantum dots in 2D materials \cite{Pisoni2017, Zhang2017, Pisoni2018, Wang2018, Davari2020}. Gated structures were used to create many-electron single and double quantum dots in eight layers of molybdenum disulfide ($\mathrm{MoS_2}$)\cite{Zhang2017}, in monolayer $\mathrm{MoS_2}$\cite{Pisoni2018}, as well as in monolayer and bilayer tungsten diselenide ($\mathrm{WSe_2}$)\cite{Davari2020}. In these low dimensional structures,defects and disorder affect the transport properties of real devices, where they alter the electrostatic environment experienced by the charge carriers
 \cite{rhodes2019disorder, plumadore2020defects}. Effects of such a potential landscape have been seen in experiments reporting resonant transport through quantum dots that form in local minima \cite{Guo1.2015, Guo2.2015}.
These experiments underscore not only the promise of the  2D materials platform for quantum devices, but also the remaining challenges in material purity, device fabrication and ultimate electrical and optical control over the charge, spin and valleys degrees of freedom of the quantum states. Here, we address some of these questions and present measurements of hole transport through an electrostatically controlled quantum dot in monolayer $\mathrm{WSe_2}$ at a temperature of 4K.

The $\mathrm{WSe_2}$ device was assembled using standard dry transfer methods \cite{Wang2013,Boddison2019assembly,Boddison2019flat} on a  $\mathrm{Si/SiO_2}$ wafer with 285 nm $\mathrm{SiO_2}$ and degenerately doped Si. A schematic representation and an optical micrograph of a completed device are shown in Figure 1. Hexagonal boron nitride (hBN) (38nm), used as an insulating dielectric, was transferred on top of pre-patterned local control gates (Ti (5 nm) / Au (5 nm)), followed by patterned bottom electrical contacts (Cr (2 nm) / Pt (8 nm))\cite{Movva2015} using e-beam lithography. To minimize the contact resistance, we mechanically removed the polymer residues using an atomic force microscope tip in contact mode\cite{Goossens2012,Rosenberger2018}. Following this key step, we used dry transfer techniques to position a $\mathrm{WSe_2}$ monolayer encapsulated by a hBN flake ($\mathrm{\approx30}$ nm) onto the mechanically cleaned bottom contacts. In a final lithographic step, we deposited a top gate (Ti (5 nm) / Pd (20 nm) / Au (100 nm)), which covered the entire $\mathrm{WSe_2}$ flake. This gate was used to modify the charge carrier density in the $\mathrm{WSe_2}$ and to achieve a regime of ohmic contact. 

\begin{figure*}
    \centering
    \includegraphics[width=\textwidth]{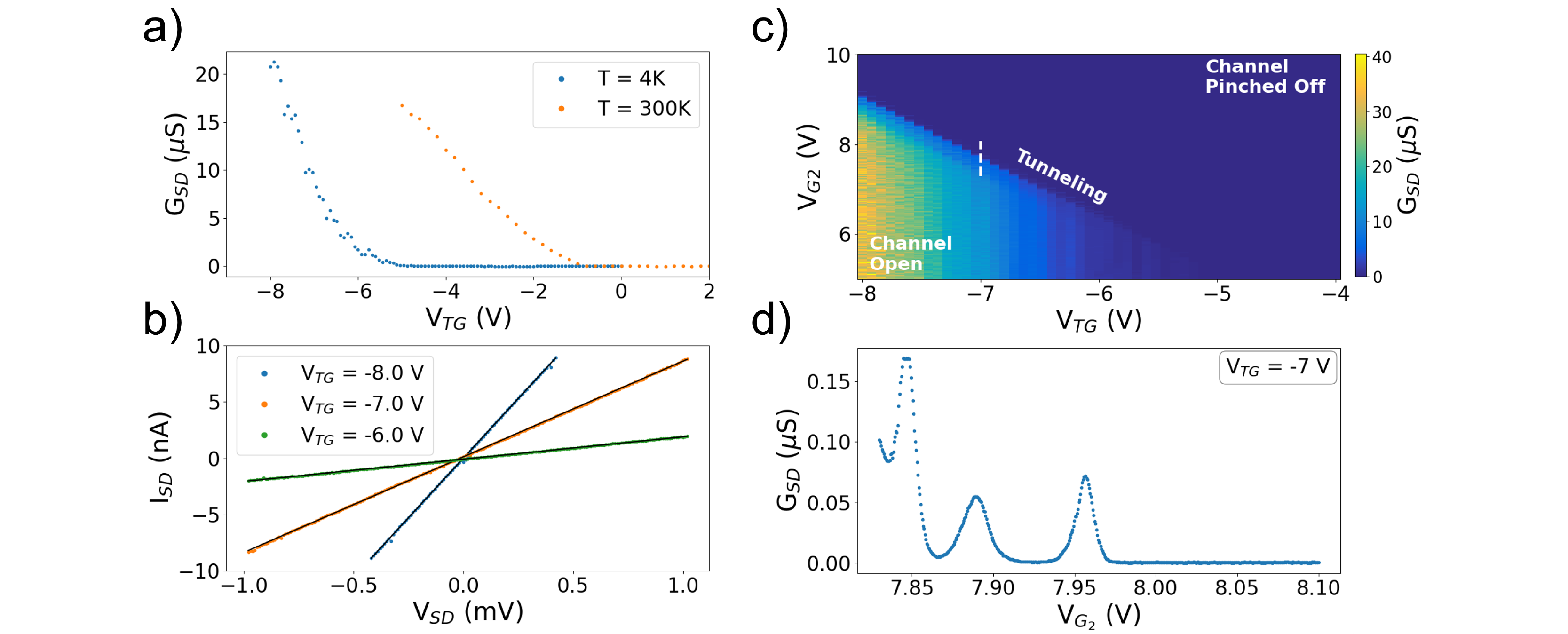}
    \caption{(a) Source-drain conductance as a function of top gate voltage where the black (blue) curve was taken at room temperature (4K). (b) Source-drain current versus source-drain voltage taken at various top gate voltages, while all control gates are grounded. (c) Source-drain conductance versus top gate and control gates $\mathrm{V_{G2}}$ (upper-middle and lower-middle in inset 1b). The conductance across the channel operates in three distinct regimes: an open channel, a tunneling regime, and a pinched-off regime. (d) Source-drain conductance versus the control gate voltage $\mathrm{V_{G2}}$ at a constant top gate voltage of -7 V.}
    \label{fig:fig2}
\end{figure*}
Experimental characterisation of this device was initially performed by applying a small DC voltage to the source contact and monitoring the drain current, while sweeping a DC voltage on the top gate (Figure 2a). The increase in current for negative top gate voltages indicates the successful population of holes in the channel, consistent with the doping levels found in previous experiments \cite{Movva2015, Davari2020}. Furthermore, as the temperature is lowered from room temperature to 4K, a larger negative voltage on the top gate is necessary to activate the conducting channel, as the microscopic crystal defects decrease the device mobility especially at low temperatures \cite{pradhan2015hall}. The remaining measurements shown in this letter are performed at 4K. Figure 2b shows current-voltage characteristics obtained at different values of top gate voltage, as indicated. We note that in our device ohmic contacts are achieved when the top gate is tuned below -5 V, as evidenced by the linear behaviour of the $\mathrm{I_{SD}-V_{SD}}$ curves. A total resistance of 47 $\mathrm{k\Omega}$ is observed at a top gate voltage of $\mathrm{V_{TG}}$=-8 V.

Next, we investigated the influence of the local control gates on the transport proprieties of the device, by applying identical voltages to pairs of control gates (upper (U) and lower(L)), while keeping the remaining control gates grounded. Figure 2c shows the conductivity $\mathrm{G_{SD}}$ across the $\mathrm{WSe_2}$ channel as a function of top gate voltage $\mathrm{V_{TG}}$ and the voltage applied to both control gates in the pair $\mathrm{G_{2U}}$ and $\mathrm{G_{2L}}$ simultaneously, $\mathrm{V_{G2}}$. We distinguish three regimes. When the channel is activated by the top gate and the control gate voltages are below a certain threshold, the channel is open (conductive), and a finite current is measured. As we increase the local control gate voltage $\mathrm{V_{G2}}$, the channel crosses over to a resonant tunneling regime and is eventually completely pinched-off with sufficiently large control gate voltages. For clarity, Figure 2d shows a line trace along the range of voltages indicated by the dashed line in Figure 2c, taken at a top gate voltage $\mathrm{V_{TG}}$ = -7 V. The presence of resonant transport peaks in the tunneling regime suggest the presence of quantum dots in the device.

We can further explore the characteristics of the quantum dots in the channel by varying the voltages on the local control gates. Figure 3a shows the changes in the source-drain current $\mathrm{I_{SD}}$, when we vary the voltages on $\mathrm{G_{2U}}$ and $\mathrm{G_{2L}}$, at a fixed top gate voltage $\mathrm{V_{TG}=-5.5V}$. In this charge stability diagram, we identify horizontal lines and vertical lines, corresponding to different transport resonances through quantum dots. Vertical lines indicate that the quantum dot is capacitively coupled only to gate $\mathrm{G_{2U}}$ and, therefore, that its location is under the gate $\mathrm{G_{2U}}$ or very close to it. Similarly, horizontal peaks are entirely controlled by gate $\mathrm{G_{2L}}$. We can also identify quantum dots formed in between gates as depicted by the $\mathrm{G_{3L}}$ - $\mathrm{G_{3U}}$ stability diagram in Figure 3b, where in addition to having horizontal and vertical lines, we find diagonal lines with various slopes, indicating the presence
of quantum dots forming at various locations  between the two gates. In this situation, we observe avoided crossings as the one highlighted by the red curves in Figure 3b, suggesting capacitive coupling between such quantum dots.
The results presented in both Figure 3a and Figure 3b imply that the potential landscape of the $\mathrm{WSe_2}$ varies across the sample, creating local quantum dots that are identified by transport resonances near depletion. The control gates can therefore be used to study individual quantum dots.

\begin{figure*}
    \centering
    \includegraphics[width=\textwidth]{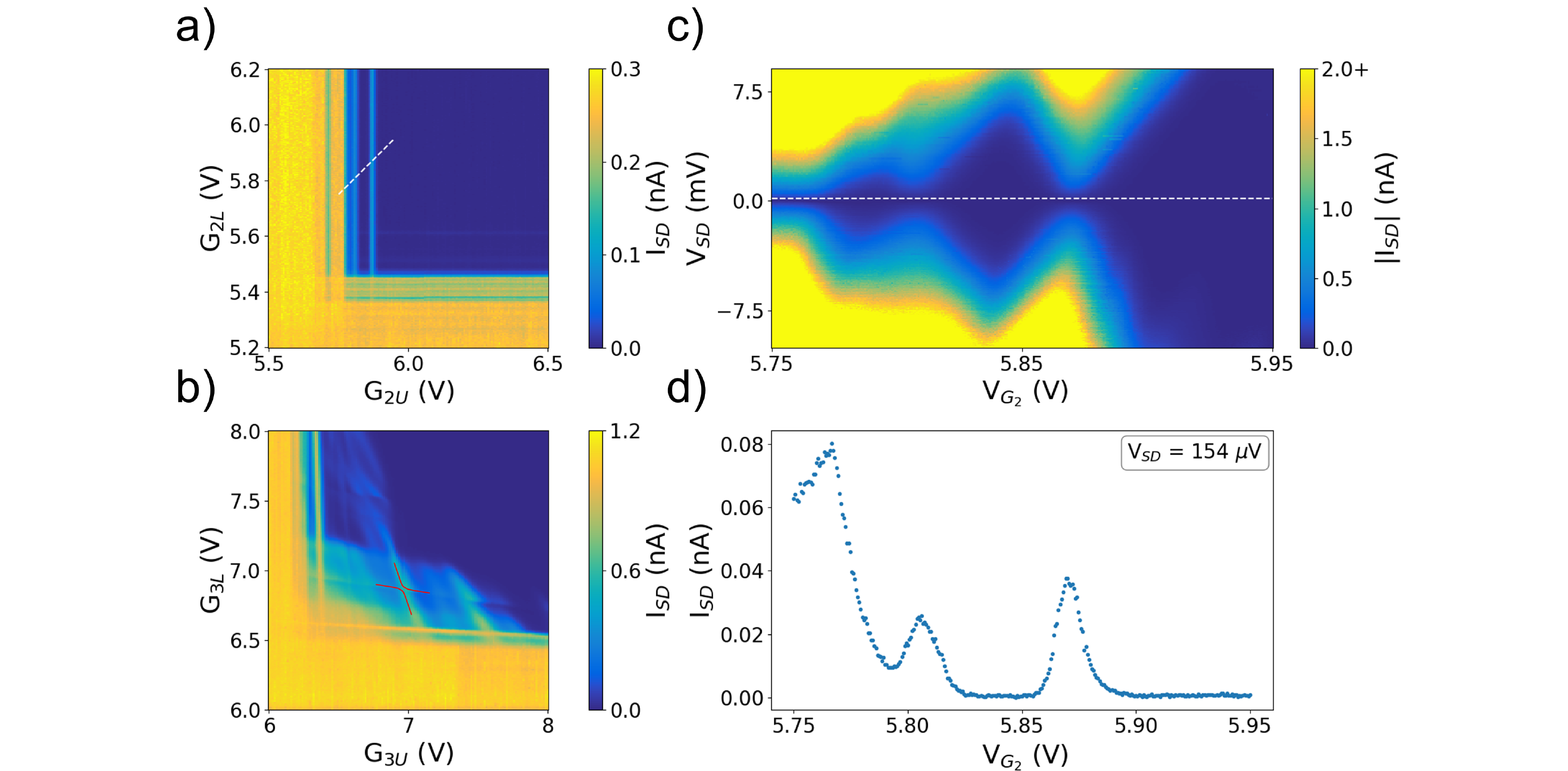}
    \caption{(a) Source-drain current $\mathrm{I_{SD}}$ as a function of individual control gates $\mathrm{G_{2L}}$ and $\mathrm{G_{2U}}$ taken at $\mathrm{V_{TG} = -5.5}$ V and $\mathrm{V_{SD} = 0.5}$ mV. (b) Source-drain current $\mathrm{I_{SD}}$ as a function of two other individual control gates $\mathrm{G_{3L}}$ and $\mathrm{G_{3U}}$ taken at $\mathrm{V_{TG} = -6.0}$ V and $\mathrm{V_{SD} = 0.5}$ mV. (c) Source-drain current $\mathrm{I_{SD}}$ as a function of bias voltage $\mathrm{V_{SD}}$ and control gate $\mathrm{V_{G2}}$ voltage corresponding to the white dashed line in (a). (d) Source-drain current $\mathrm{I_{SD}}$ as a function of $\mathrm{G_{2L\&R}}$ taken at a $\mathrm{V_{SD} = 154}$ $\mathrm{\mu}V$ corresponding to the white dashed line in (c).}
    \label{fig:fig3}
\end{figure*}

We explore the properties of the quantum dot associated with the vertical lines in Figure 3a, by sweeping the bias voltage $\mathrm{V_{SD}}$ for different values of $\mathrm{G_{2L}}$ and $\mathrm{G_{2U}}$ as indicated by the dashed line, while the top gate voltage is kept constant $\mathrm{V_{TG}}=-5.5$ V.  The variation of the source-drain current $\mathrm{I_{SD}}$ while the bias and control gate voltages are varied, is shown in Figure 3c.  We note the presence of diamond-shaped features (Coulomb diamonds) characteristic of quantum dot behaviour, where regions with zero source-drain current represent a regime of Coulomb blockaded transport. From the plot of Figure 3c we can estimate the size and charging energy of the quantum dot. From the height of the first Coulomb diamond, the charging energy $\mathrm{U \sim3.4}$ meV, value consistent with previous reports in gated QD structures in $\mathrm{WSe_2}$ \cite{Davari2020,Guo1.2015}. Figure 3d shows a line cut through the map of Figure 3c, taken at a source-drain voltage $\mathrm{V_{SD}=154}$ $\mathrm{\mu}V$. The energy separation between the peaks is $\mathrm{{\Delta}V_G\approx65}$ mV. We can estimate the size of the quantum dot by approximating the quantum dot and the control gate as a parallel plate capacitor. Using $\mathrm{\epsilon_{hBN}=4}$ as hBN’s relative permittivity\cite{Young2012} and its thickness $\mathrm{t_{hBN}=38}$ nm, the gate capacitance is approximated to be $\mathrm{C_G=\epsilon_0 \epsilon_{hBN} {\pi} d^2/4t_{hBN}}$, where $\mathrm{d}$ is the quantum dot diameter. The gate capacitance can also be expressed as $\mathrm{C_G=e^2/{{\Delta}V_G}}$. From these two equations, we estimate the quantum dot diameter to be $\mathrm{d \approx 58}$ nm. 


To understand the nature of electronic states available for holes that are confined in a  $\mathrm{WSe_2}$ quantum dot as well as the interplay of valley, spin and orbital degrees of freedom, we performed tight binding  and ab-initio calculations (using ABINIT) of valence holes in single layer $\mathrm{WSe_2}$ \cite{bieniek_hawrylak_prb2018,altintas_hawrylak_xxx2021}  and confined to a gated quantum dot  \cite{bieniek_hawrylak_prb2020,szulakowska_hawrylak_prb2020, altintas_hawrylak_xxx2021}. A single layer of $\mathrm{WSe_2}$ consists of two triangular sublattices: sublattice A, having d orbitals from W atoms and, sublattice B, consisting of a pair of Se atoms, with $p$-orbitals below and above the metal layer. A $\mathrm{WSe_2}$ monolayer with such a hexagonal lattice is  a direct gap semiconductor with energy minima at the two nonequivalent valleys in the Brillouin zone. Figure 4a shows the calculated energy spectrum of holes in a single layer of $\mathrm{WSe_2}$ around the two non-equivalent valleys, $+K$ and $-K$, where  strong spin-orbit interaction leads to splitting of hole spin up and spin down states by ~$470$ meV. This implies that the spin degree of freedom of the hole ground states is locked to the valley index with spin up locked to $+K$ and spin down locked to $-K$. In addition, the hole wavefunction changes its orbital character, at $+K$ valley it is built of $m=+2$ $W$ d-orbital, while at $-K$ valley it is built of $m=-2$ W d-orbital. The energy bands reflect the massive Dirac fermion description of the energy bands in $\mathrm{WSe_2}$, where close to the top of the upper valence band the dispersion is approximately parabolic with effective mass $m^*=0.28$, but it  quickly becomes linear. 

\begin{figure*}
    \centering
    \includegraphics[width=\textwidth]{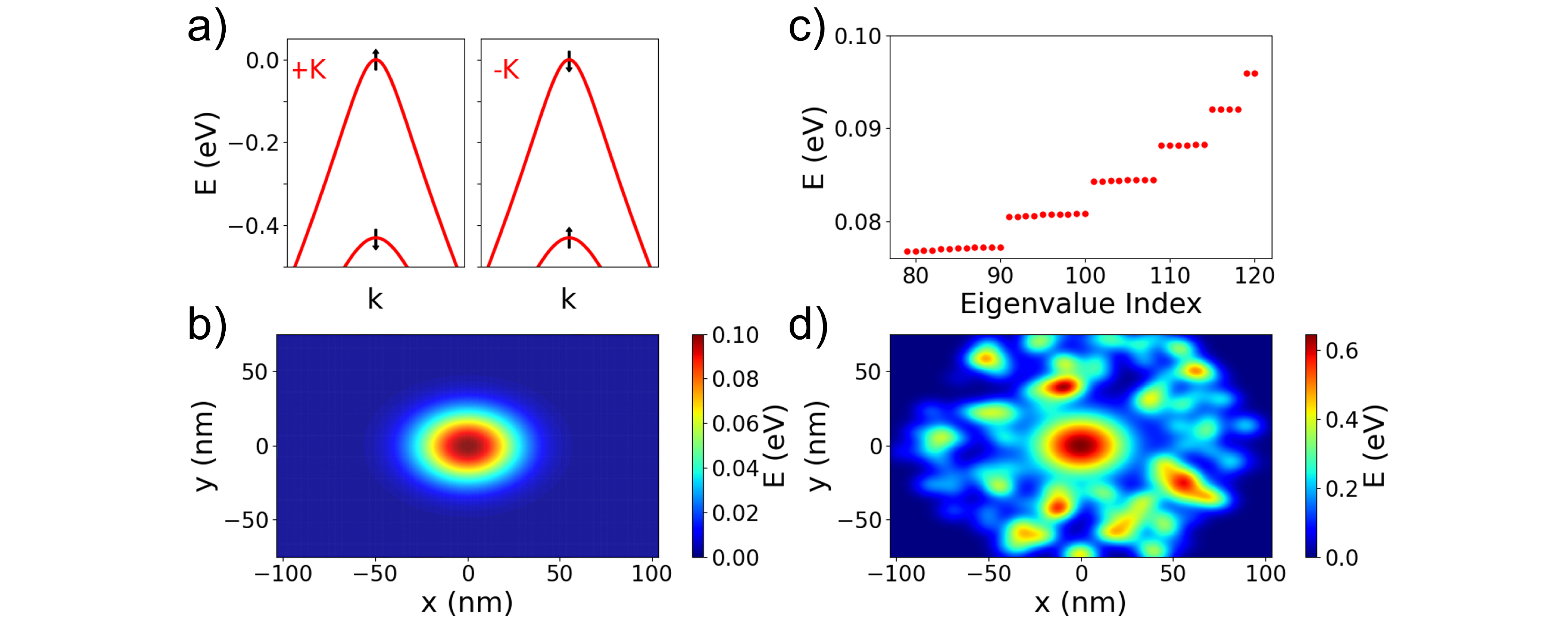}
    \caption{(a) Electronic band dispersion for $WSe_2$ around the top of the valence band for the two valleys $+K$ and $-K$, emphasizing the spin structure. (b) Schematic view of a simulated Gaussian potential for a quantum dot in $WSe_2$. (c) Energy spectrum of hole states in a  $WSe_2$ quantum dot confined by the  potential shown in (b). 
    (d) Schematic representation of a quantum dot generated  by a gate and surrounded by a random electrostatic potential.}

    \label{fig:fig4}
\end{figure*}

We consider hole states localised in a $\mathrm{WSe_2}$ quantum dot, similar to those leading to the observed Coulomb blockade spectrum in our device. Figure 4b represents an idealized Gaussian potential profile generated by an external gate  \cite{szulakowska_hawrylak_prb2020,altintas_hawrylak_xxx2021} with $V_0=100$ meV and diameter of $60$ nm, assumed to confine hole states in a  $\mathrm{WSe_2}$ layer. The calculated spectrum for this quantum dot is shown in Figure 4c, where the zero energy is at the top of the valence band and the confining potential leads to discrete quantum dot levels in the energy gap of the $\mathrm{WSe_2}$ layer. As a result of confinement, we find several energy shells, separated by energy $\omega_0=5$ meV. The lowest hole energy s-shell is doubly degenerate, corresponding to $(+K,\uparrow,m=+2)$ and $(-K,\downarrow,m=-2)$ states. The next energy p-shell is almost fourfold degenerate, followed by a six fold degenerate d-shell.
The predicted hole energy spectrum is similar to the energy spectrum of carriers confined to semiconductor parabolic quantum dots\cite{raymond_hawrylak_prl2004}, except for lifting the degeneracy of hole shells by topological magnetic moments inherent to massive Dirac fermions \cite{bieniek_hawrylak_prb2020,szulakowska_hawrylak_prb2020, altintas_hawrylak_xxx2021}.
To estimate the charging energy $U$ needed to add a hole to the lowest state, we consider a parabolic quantum dot, so that $U=Ry^* \sqrt{\pi} \sqrt{\omega_0/Ry^{*}}$ \cite{sheng_hawrylak_pss2003}, where $Ry^*$ is an effective Rydberg, given by $Ry^*= Ry \times m^{*}/(\epsilon^2)$, where $m^{*}$ is the effective mass  and $\epsilon$ the effective dielectric constant. For typical parameters $\epsilon = 10$ and $m^*=0.28$ characteristic of holes in $\mathrm{WSe_2}$, the resulting charging energy is found to be $U=20$ meV. We note that the discrepancy between this simple model and the measured experimental results are likely due to the presence of random electrostatic potential below the semiconducting channel or the gate, as schematically illustrated in Figure 4d. This suggests that to fully take advantage of the unique properties of quantum confined states in 2D semiconductors, more effort is needed to optimize the materials homogeneity and purity, as well as to improve electrical contacts.

In summary, we demonstrated electrical control of $\mathrm{WSe_2}$ hole quantum dots in gated devices based on 2D heterostructures. Electrical transport measurements revealed a Coulomb blockade regime, from which we estimated the charging energy and size of the formed quantum dots. Additionally, charge stability diagrams using control gates indicate that the quantum dots are likely generated by the inherent electrostatic fluctuations of the device and that they can be coupled. Our theoretical calculations reveal that the ground state of holes in such a quantum dot is not fourfold degenerate as expected due to the spin and valley degrees of freedom, but doubly degenerate due to valley-spin locking. These results shed light on mesoscopic transport features in 2D semiconducting quantum dot devices, paving the way for the development of quantum devices based on electrostatically confined $\mathrm{WSe_2}$ quantum dots.


\begin{acknowledgments}
The authors thank  A. Dusko and Y.Saleem for useful discussions.
This work was supported by the High Throughput and Secure Networks Challenge Program at the National Research Council of Canada.
AL-M and JBC acknowledge funding from the National Sciences and Engineering Research Council (NSERC) Discovery Grant RGPIN-2016-06717.  We also acknowledge the support of the Natural Sciences and Engineering Research Council of Canada (NSERC) through Strategic Project STPGP 521420. PH acknowledges NSERC Discovery grant and University of Ottawa Research Chair in Quantum Theory of Materials, Nanostructures and Devices. M.B acknowledges financial support from Polish National Agency for Academic Exchange (NAWA), Poland, grant PPI/APM/2019/1/00085/U/00001. Computing resources from Compute Canada are gratefully acknowledged.
K.W. and T.T. acknowledge support from the Elemental Strategy Initiative conducted by the MEXT, Japan (Grant Number JPMXP0112101001) and  JSPS KAKENHI (Grant Numbers 19H05790 and JP20H00354).

\end{acknowledgments}

\section*{Data Availability}

The data that support the findings of this study are available from the corresponding author upon reasonable request.

\bibliography{main}

\providecommand{\noopsort}[1]{}\providecommand{\singleletter}[1]{#1}%
\begin{thebibliography}{39}%
\makeatletter
\providecommand \@ifxundefined [1]{%
 \@ifx{#1\undefined}
}%
\providecommand \@ifnum [1]{%
 \ifnum #1\expandafter \@firstoftwo
 \else \expandafter \@secondoftwo
 \fi
}%
\providecommand \@ifx [1]{%
 \ifx #1\expandafter \@firstoftwo
 \else \expandafter \@secondoftwo
 \fi
}%
\providecommand \natexlab [1]{#1}%
\providecommand \enquote  [1]{``#1''}%
\providecommand \bibnamefont  [1]{#1}%
\providecommand \bibfnamefont [1]{#1}%
\providecommand \citenamefont [1]{#1}%
\providecommand \href@noop [0]{\@secondoftwo}%
\providecommand \href [0]{\begingroup \@sanitize@url \@href}%
\providecommand \@href[1]{\@@startlink{#1}\@@href}%
\providecommand \@@href[1]{\endgroup#1\@@endlink}%
\providecommand \@sanitize@url [0]{\catcode `\\12\catcode `\$12\catcode
  `\&12\catcode `\#12\catcode `\^12\catcode `\_12\catcode `\%12\relax}%
\providecommand \@@startlink[1]{}%
\providecommand \@@endlink[0]{}%
\providecommand \url  [0]{\begingroup\@sanitize@url \@url }%
\providecommand \@url [1]{\endgroup\@href {#1}{\urlprefix }}%
\providecommand \urlprefix  [0]{URL }%
\providecommand \Eprint [0]{\href }%
\providecommand \doibase [0]{http://dx.doi.org/}%
\providecommand \selectlanguage [0]{\@gobble}%
\providecommand \bibinfo  [0]{\@secondoftwo}%
\providecommand \bibfield  [0]{\@secondoftwo}%
\providecommand \translation [1]{[#1]}%
\providecommand \BibitemOpen [0]{}%
\providecommand \bibitemStop [0]{}%
\providecommand \bibitemNoStop [0]{.\EOS\space}%
\providecommand \EOS [0]{\spacefactor3000\relax}%
\providecommand \BibitemShut  [1]{\csname bibitem#1\endcsname}%
\let\auto@bib@innerbib\@empty
\bibitem [{\citenamefont {Brum}\ and\ \citenamefont
  {Hawrylak}(1997)}]{brum_hawrylak_SLM1997}%
  \BibitemOpen
  \bibfield  {author} {\bibinfo {author} {\bibfnamefont {J.~A.}\ \bibnamefont
  {Brum}}\ and\ \bibinfo {author} {\bibfnamefont {P.}~\bibnamefont
  {Hawrylak}},\ }\href@noop {} {\bibfield  {journal} {\bibinfo  {journal}
  {Superlattices and Microstructures}\ }\textbf {\bibinfo {volume} {22}},\
  \bibinfo {pages} {431} (\bibinfo {year} {1997})}\BibitemShut {NoStop}%
\bibitem [{\citenamefont {Loss}\ and\ \citenamefont
  {DiVincenzo}(1998)}]{Loss1998}%
  \BibitemOpen
  \bibfield  {author} {\bibinfo {author} {\bibfnamefont {D.}~\bibnamefont
  {Loss}}\ and\ \bibinfo {author} {\bibfnamefont {D.~P.}\ \bibnamefont
  {DiVincenzo}},\ }\href@noop {} {\bibfield  {journal} {\bibinfo  {journal}
  {Physical Review A}\ }\textbf {\bibinfo {volume} {57}} (\bibinfo {year}
  {1998})}\BibitemShut {NoStop}%
\bibitem [{\citenamefont {Ciorga}\ \emph {et~al.}(2000)\citenamefont {Ciorga},
  \citenamefont {Sachrajda}, \citenamefont {Hawrylak}, \citenamefont {Gould},
  \citenamefont {Zawadzki}, \citenamefont {Jullian}, \citenamefont {Feng},\
  and\ \citenamefont {Wasilewski}}]{Ciorga2000}%
  \BibitemOpen
  \bibfield  {author} {\bibinfo {author} {\bibfnamefont {M.}~\bibnamefont
  {Ciorga}}, \bibinfo {author} {\bibfnamefont {A.~S.}\ \bibnamefont
  {Sachrajda}}, \bibinfo {author} {\bibfnamefont {P.}~\bibnamefont {Hawrylak}},
  \bibinfo {author} {\bibfnamefont {C.}~\bibnamefont {Gould}}, \bibinfo
  {author} {\bibfnamefont {P.}~\bibnamefont {Zawadzki}}, \bibinfo {author}
  {\bibfnamefont {S.}~\bibnamefont {Jullian}}, \bibinfo {author} {\bibfnamefont
  {Y.}~\bibnamefont {Feng}}, \ and\ \bibinfo {author} {\bibfnamefont
  {Z.}~\bibnamefont {Wasilewski}},\ }\href@noop {} {\bibfield  {journal}
  {\bibinfo  {journal} {Physical Review B}\ }\textbf {\bibinfo {volume} {61}}
  (\bibinfo {year} {2000})}\BibitemShut {NoStop}%
\bibitem [{\citenamefont {Elzerman}\ \emph {et~al.}(2003)\citenamefont
  {Elzerman}, \citenamefont {Hanson}, \citenamefont {Greidanus}, \citenamefont
  {van Beveren}, \citenamefont {Franceschi}, \citenamefont {Vandersypen},
  \citenamefont {Tarucha},\ and\ \citenamefont {Kouwenhoven}}]{Elzerman2003}%
  \BibitemOpen
  \bibfield  {author} {\bibinfo {author} {\bibfnamefont {J.~M.}\ \bibnamefont
  {Elzerman}}, \bibinfo {author} {\bibfnamefont {R.}~\bibnamefont {Hanson}},
  \bibinfo {author} {\bibfnamefont {J.~S.}\ \bibnamefont {Greidanus}}, \bibinfo
  {author} {\bibfnamefont {L.~H.~W.}\ \bibnamefont {van Beveren}}, \bibinfo
  {author} {\bibfnamefont {S.~D.}\ \bibnamefont {Franceschi}}, \bibinfo
  {author} {\bibfnamefont {L.~M.~K.}\ \bibnamefont {Vandersypen}}, \bibinfo
  {author} {\bibfnamefont {S.}~\bibnamefont {Tarucha}}, \ and\ \bibinfo
  {author} {\bibfnamefont {L.~P.}\ \bibnamefont {Kouwenhoven}},\ }\href@noop {}
  {\bibfield  {journal} {\bibinfo  {journal} {Physical Review B}\ }\textbf
  {\bibinfo {volume} {67}} (\bibinfo {year} {2003})}\BibitemShut {NoStop}%
\bibitem [{\citenamefont {Gaudreau}\ \emph {et~al.}(2009)\citenamefont
  {Gaudreau}, \citenamefont {Kam}, \citenamefont {Granger}, \citenamefont
  {Studenikin}, \citenamefont {Zawadzki},\ and\ \citenamefont
  {Sachrajda}}]{Gaudreau2009}%
  \BibitemOpen
  \bibfield  {author} {\bibinfo {author} {\bibfnamefont {L.}~\bibnamefont
  {Gaudreau}}, \bibinfo {author} {\bibfnamefont {A.}~\bibnamefont {Kam}},
  \bibinfo {author} {\bibfnamefont {G.}~\bibnamefont {Granger}}, \bibinfo
  {author} {\bibfnamefont {S.~A.}\ \bibnamefont {Studenikin}}, \bibinfo
  {author} {\bibfnamefont {P.}~\bibnamefont {Zawadzki}}, \ and\ \bibinfo
  {author} {\bibfnamefont {A.~S.}\ \bibnamefont {Sachrajda}},\ }\href@noop {}
  {\bibfield  {journal} {\bibinfo  {journal} {Applied Physics Letters}\
  }\textbf {\bibinfo {volume} {95}} (\bibinfo {year} {2009})}\BibitemShut
  {NoStop}%
\bibitem [{\citenamefont {Zajac}\ \emph {et~al.}(2016)\citenamefont {Zajac},
  \citenamefont {Hazard}, \citenamefont {Mi}, \citenamefont {Nielsen},\ and\
  \citenamefont {Petta}}]{Zajac2016}%
  \BibitemOpen
  \bibfield  {author} {\bibinfo {author} {\bibfnamefont {D.}~\bibnamefont
  {Zajac}}, \bibinfo {author} {\bibfnamefont {T.}~\bibnamefont {Hazard}},
  \bibinfo {author} {\bibfnamefont {X.}~\bibnamefont {Mi}}, \bibinfo {author}
  {\bibfnamefont {E.}~\bibnamefont {Nielsen}}, \ and\ \bibinfo {author}
  {\bibfnamefont {J.}~\bibnamefont {Petta}},\ }\href@noop {} {\bibfield
  {journal} {\bibinfo  {journal} {Physical Review Applied}\ }\textbf {\bibinfo
  {volume} {6}} (\bibinfo {year} {2016})}\BibitemShut {NoStop}%
\bibitem [{\citenamefont {Nichol}\ \emph {et~al.}(2017)\citenamefont {Nichol},
  \citenamefont {Orona}, \citenamefont {Harvey}, \citenamefont {Fallahi},
  \citenamefont {Gardner}, \citenamefont {Manfra},\ and\ \citenamefont
  {Yacoby}}]{Nichol2017}%
  \BibitemOpen
  \bibfield  {author} {\bibinfo {author} {\bibfnamefont {J.~M.}\ \bibnamefont
  {Nichol}}, \bibinfo {author} {\bibfnamefont {L.~A.}\ \bibnamefont {Orona}},
  \bibinfo {author} {\bibfnamefont {S.~P.}\ \bibnamefont {Harvey}}, \bibinfo
  {author} {\bibfnamefont {S.}~\bibnamefont {Fallahi}}, \bibinfo {author}
  {\bibfnamefont {G.~C.}\ \bibnamefont {Gardner}}, \bibinfo {author}
  {\bibfnamefont {M.~J.}\ \bibnamefont {Manfra}}, \ and\ \bibinfo {author}
  {\bibfnamefont {A.}~\bibnamefont {Yacoby}},\ }\href {www.nature.com/npjqi}
  {\bibfield  {journal} {\bibinfo  {journal} {npj Quantum Information}\
  }\textbf {\bibinfo {volume} {3}},\ \bibinfo {pages} {3} (\bibinfo {year}
  {2017})}\BibitemShut {NoStop}%
\bibitem [{\citenamefont {Zajac}\ \emph {et~al.}(2018)\citenamefont {Zajac},
  \citenamefont {Sigillito}, \citenamefont {Russ}, \citenamefont {Borjans},
  \citenamefont {Taylor}, \citenamefont {Burkard},\ and\ \citenamefont
  {Petta}}]{Zajac2018}%
  \BibitemOpen
  \bibfield  {author} {\bibinfo {author} {\bibfnamefont {D.~M.}\ \bibnamefont
  {Zajac}}, \bibinfo {author} {\bibfnamefont {A.~J.}\ \bibnamefont
  {Sigillito}}, \bibinfo {author} {\bibfnamefont {M.}~\bibnamefont {Russ}},
  \bibinfo {author} {\bibfnamefont {F.}~\bibnamefont {Borjans}}, \bibinfo
  {author} {\bibfnamefont {J.~M.}\ \bibnamefont {Taylor}}, \bibinfo {author}
  {\bibfnamefont {G.}~\bibnamefont {Burkard}}, \ and\ \bibinfo {author}
  {\bibfnamefont {J.~R.}\ \bibnamefont {Petta}},\ }\href@noop {} {\bibfield
  {journal} {\bibinfo  {journal} {Science}\ }\textbf {\bibinfo {volume}
  {359}},\ \bibinfo {pages} {439} (\bibinfo {year} {2018})}\BibitemShut
  {NoStop}%
\bibitem [{\citenamefont {Hanson}\ \emph {et~al.}(2007)\citenamefont {Hanson},
  \citenamefont {Kouwenhoven}, \citenamefont {Petta}, \citenamefont {Tarucha},\
  and\ \citenamefont {Vandersypen}}]{Hanson2007RMP}%
  \BibitemOpen
  \bibfield  {author} {\bibinfo {author} {\bibfnamefont {R.}~\bibnamefont
  {Hanson}}, \bibinfo {author} {\bibfnamefont {L.~P.}\ \bibnamefont
  {Kouwenhoven}}, \bibinfo {author} {\bibfnamefont {J.~R.}\ \bibnamefont
  {Petta}}, \bibinfo {author} {\bibfnamefont {S.}~\bibnamefont {Tarucha}}, \
  and\ \bibinfo {author} {\bibfnamefont {L.~M.~K.}\ \bibnamefont
  {Vandersypen}},\ }\href {\doibase 10.1103/RevModPhys.79.1217} {\bibfield
  {journal} {\bibinfo  {journal} {Rev. Mod. Phys.}\ }\textbf {\bibinfo {volume}
  {79}},\ \bibinfo {pages} {1217} (\bibinfo {year} {2007})}\BibitemShut
  {NoStop}%
\bibitem [{\citenamefont {Veldhorst}\ \emph {et~al.}(2014)\citenamefont
  {Veldhorst}, \citenamefont {Hwang}, \citenamefont {Yang}, \citenamefont
  {Leenstra}, \citenamefont {de~Ronde}, \citenamefont {Dehollain},
  \citenamefont {Muhonen}, \citenamefont {Hudson}, \citenamefont {Itoh},
  \citenamefont {Morello},\ and\ \citenamefont {Dzurak}}]{Veldhorst2014}%
  \BibitemOpen
  \bibfield  {author} {\bibinfo {author} {\bibfnamefont {M.}~\bibnamefont
  {Veldhorst}}, \bibinfo {author} {\bibfnamefont {J.~C.~C.}\ \bibnamefont
  {Hwang}}, \bibinfo {author} {\bibfnamefont {C.~H.}\ \bibnamefont {Yang}},
  \bibinfo {author} {\bibfnamefont {A.~W.}\ \bibnamefont {Leenstra}}, \bibinfo
  {author} {\bibfnamefont {B.}~\bibnamefont {de~Ronde}}, \bibinfo {author}
  {\bibfnamefont {J.~P.}\ \bibnamefont {Dehollain}}, \bibinfo {author}
  {\bibfnamefont {J.~T.}\ \bibnamefont {Muhonen}}, \bibinfo {author}
  {\bibfnamefont {F.~E.}\ \bibnamefont {Hudson}}, \bibinfo {author}
  {\bibfnamefont {K.~M.}\ \bibnamefont {Itoh}}, \bibinfo {author}
  {\bibfnamefont {A.}~\bibnamefont {Morello}}, \ and\ \bibinfo {author}
  {\bibfnamefont {A.~S.}\ \bibnamefont {Dzurak}},\ }\href@noop {} {\bibfield
  {journal} {\bibinfo  {journal} {Nature Nanotechnology}\ }\textbf {\bibinfo
  {volume} {9}},\ \bibinfo {pages} {981} (\bibinfo {year} {2014})}\BibitemShut
  {NoStop}%
\bibitem [{\citenamefont {Lodahl}(2017)}]{Lodahl2017}%
  \BibitemOpen
  \bibfield  {author} {\bibinfo {author} {\bibfnamefont {P.}~\bibnamefont
  {Lodahl}},\ }\href {\doibase 10.1088/2058-9565/aa91bb} {\bibfield  {journal}
  {\bibinfo  {journal} {Quantum Science and Technology}\ }\textbf {\bibinfo
  {volume} {3}},\ \bibinfo {pages} {013001} (\bibinfo {year}
  {2017})}\BibitemShut {NoStop}%
\bibitem [{\citenamefont {Geim}\ and\ \citenamefont
  {Grigorieva}(2013)}]{Geim2013}%
  \BibitemOpen
  \bibfield  {author} {\bibinfo {author} {\bibfnamefont {A.~K.}\ \bibnamefont
  {Geim}}\ and\ \bibinfo {author} {\bibfnamefont {I.~V.}\ \bibnamefont
  {Grigorieva}},\ }\href@noop {} {\bibfield  {journal} {\bibinfo  {journal}
  {Nature}\ }\textbf {\bibinfo {volume} {499}},\ \bibinfo {pages} {419}
  (\bibinfo {year} {2013})}\BibitemShut {NoStop}%
\bibitem [{\citenamefont {Atat{\"u}re}\ \emph {et~al.}(2018)\citenamefont
  {Atat{\"u}re}, \citenamefont {Englund}, \citenamefont {Vamivakas},
  \citenamefont {Lee},\ and\ \citenamefont {Wrachtrup}}]{atature2018material}%
  \BibitemOpen
  \bibfield  {author} {\bibinfo {author} {\bibfnamefont {M.}~\bibnamefont
  {Atat{\"u}re}}, \bibinfo {author} {\bibfnamefont {D.}~\bibnamefont
  {Englund}}, \bibinfo {author} {\bibfnamefont {N.}~\bibnamefont {Vamivakas}},
  \bibinfo {author} {\bibfnamefont {S.-Y.}\ \bibnamefont {Lee}}, \ and\
  \bibinfo {author} {\bibfnamefont {J.}~\bibnamefont {Wrachtrup}},\ }\href@noop
  {} {\bibfield  {journal} {\bibinfo  {journal} {Nature Reviews Materials}\
  }\textbf {\bibinfo {volume} {3}},\ \bibinfo {pages} {38} (\bibinfo {year}
  {2018})}\BibitemShut {NoStop}%
\bibitem [{\citenamefont {Liu}\ and\ \citenamefont
  {Hersam}(2019)}]{liu2019qis}%
  \BibitemOpen
  \bibfield  {author} {\bibinfo {author} {\bibfnamefont {X.}~\bibnamefont
  {Liu}}\ and\ \bibinfo {author} {\bibfnamefont {M.~C.}\ \bibnamefont
  {Hersam}},\ }\href@noop {} {\bibfield  {journal} {\bibinfo  {journal} {Nature
  Reviews Materials}\ }\textbf {\bibinfo {volume} {4}},\ \bibinfo {pages} {669}
  (\bibinfo {year} {2019})}\BibitemShut {NoStop}%
\bibitem [{\citenamefont {Mak}\ \emph {et~al.}(2010)\citenamefont {Mak},
  \citenamefont {Lee}, \citenamefont {Hone}, \citenamefont {Shan},\ and\
  \citenamefont {Heinz}}]{Mak2010}%
  \BibitemOpen
  \bibfield  {author} {\bibinfo {author} {\bibfnamefont {K.~F.}\ \bibnamefont
  {Mak}}, \bibinfo {author} {\bibfnamefont {C.}~\bibnamefont {Lee}}, \bibinfo
  {author} {\bibfnamefont {J.}~\bibnamefont {Hone}}, \bibinfo {author}
  {\bibfnamefont {J.}~\bibnamefont {Shan}}, \ and\ \bibinfo {author}
  {\bibfnamefont {T.~F.}\ \bibnamefont {Heinz}},\ }\href@noop {} {\bibfield
  {journal} {\bibinfo  {journal} {Physical Review Letters}\ }\textbf {\bibinfo
  {volume} {105}} (\bibinfo {year} {2010})}\BibitemShut {NoStop}%
\bibitem [{\citenamefont {Korm{\'a}nyos}\ \emph {et~al.}(2014)\citenamefont
  {Korm{\'a}nyos}, \citenamefont {Z{\'o}lyomi}, \citenamefont {Drummond},\ and\
  \citenamefont {Burkard}}]{kormanyos2014spin}%
  \BibitemOpen
  \bibfield  {author} {\bibinfo {author} {\bibfnamefont {A.}~\bibnamefont
  {Korm{\'a}nyos}}, \bibinfo {author} {\bibfnamefont {V.}~\bibnamefont
  {Z{\'o}lyomi}}, \bibinfo {author} {\bibfnamefont {N.~D.}\ \bibnamefont
  {Drummond}}, \ and\ \bibinfo {author} {\bibfnamefont {G.}~\bibnamefont
  {Burkard}},\ }\href@noop {} {\bibfield  {journal} {\bibinfo  {journal}
  {Physical Review X}\ }\textbf {\bibinfo {volume} {4}},\ \bibinfo {pages}
  {011034} (\bibinfo {year} {2014})}\BibitemShut {NoStop}%
\bibitem [{\citenamefont {Pisoni}\ \emph {et~al.}(2017)\citenamefont {Pisoni},
  \citenamefont {Lee}, \citenamefont {Overweg}, \citenamefont {Eich},
  \citenamefont {Simonet}, \citenamefont {Watanabe}, \citenamefont {Taniguchi},
  \citenamefont {Gorbachev}, \citenamefont {Ihn},\ and\ \citenamefont
  {Ensslin}}]{Pisoni2017}%
  \BibitemOpen
  \bibfield  {author} {\bibinfo {author} {\bibfnamefont {R.}~\bibnamefont
  {Pisoni}}, \bibinfo {author} {\bibfnamefont {Y.}~\bibnamefont {Lee}},
  \bibinfo {author} {\bibfnamefont {H.}~\bibnamefont {Overweg}}, \bibinfo
  {author} {\bibfnamefont {M.}~\bibnamefont {Eich}}, \bibinfo {author}
  {\bibfnamefont {P.}~\bibnamefont {Simonet}}, \bibinfo {author} {\bibfnamefont
  {K.}~\bibnamefont {Watanabe}}, \bibinfo {author} {\bibfnamefont
  {T.}~\bibnamefont {Taniguchi}}, \bibinfo {author} {\bibfnamefont
  {R.}~\bibnamefont {Gorbachev}}, \bibinfo {author} {\bibfnamefont
  {T.}~\bibnamefont {Ihn}}, \ and\ \bibinfo {author} {\bibfnamefont
  {K.}~\bibnamefont {Ensslin}},\ }\href@noop {} {\bibfield  {journal} {\bibinfo
   {journal} {Nano Letters}\ }\textbf {\bibinfo {volume} {17}} (\bibinfo {year}
  {2017})}\BibitemShut {NoStop}%
\bibitem [{\citenamefont {Zhang}\ \emph {et~al.}(2017)\citenamefont {Zhang},
  \citenamefont {Song}, \citenamefont {Luo}, \citenamefont {Deng},
  \citenamefont {Mosallanejad}, \citenamefont {Taniguchi}, \citenamefont
  {Watanabe}, \citenamefont {Li}, \citenamefont {Cao}, \citenamefont {Guo},
  \citenamefont {Nori},\ and\ \citenamefont {Guo}}]{Zhang2017}%
  \BibitemOpen
  \bibfield  {author} {\bibinfo {author} {\bibfnamefont {Z.-Z.}\ \bibnamefont
  {Zhang}}, \bibinfo {author} {\bibfnamefont {X.-X.}\ \bibnamefont {Song}},
  \bibinfo {author} {\bibfnamefont {G.}~\bibnamefont {Luo}}, \bibinfo {author}
  {\bibfnamefont {G.-W.}\ \bibnamefont {Deng}}, \bibinfo {author}
  {\bibfnamefont {V.}~\bibnamefont {Mosallanejad}}, \bibinfo {author}
  {\bibfnamefont {T.}~\bibnamefont {Taniguchi}}, \bibinfo {author}
  {\bibfnamefont {K.}~\bibnamefont {Watanabe}}, \bibinfo {author}
  {\bibfnamefont {H.-O.}\ \bibnamefont {Li}}, \bibinfo {author} {\bibfnamefont
  {G.}~\bibnamefont {Cao}}, \bibinfo {author} {\bibfnamefont {G.-C.}\
  \bibnamefont {Guo}}, \bibinfo {author} {\bibfnamefont {F.}~\bibnamefont
  {Nori}}, \ and\ \bibinfo {author} {\bibfnamefont {G.-P.}\ \bibnamefont
  {Guo}},\ }\href@noop {} {\bibfield  {journal} {\bibinfo  {journal} {Science
  Advances}\ }\textbf {\bibinfo {volume} {3}} (\bibinfo {year}
  {2017})}\BibitemShut {NoStop}%
\bibitem [{\citenamefont {Pisoni}\ \emph {et~al.}(2018)\citenamefont {Pisoni},
  \citenamefont {Lei}, \citenamefont {Back}, \citenamefont {Eich},
  \citenamefont {Overweg}, \citenamefont {Lee}, \citenamefont {Watanabe},
  \citenamefont {Taniguchi}, \citenamefont {Ihn},\ and\ \citenamefont
  {Ensslin}}]{Pisoni2018}%
  \BibitemOpen
  \bibfield  {author} {\bibinfo {author} {\bibfnamefont {R.}~\bibnamefont
  {Pisoni}}, \bibinfo {author} {\bibfnamefont {Z.}~\bibnamefont {Lei}},
  \bibinfo {author} {\bibfnamefont {P.}~\bibnamefont {Back}}, \bibinfo {author}
  {\bibfnamefont {M.}~\bibnamefont {Eich}}, \bibinfo {author} {\bibfnamefont
  {H.}~\bibnamefont {Overweg}}, \bibinfo {author} {\bibfnamefont
  {Y.}~\bibnamefont {Lee}}, \bibinfo {author} {\bibfnamefont {K.}~\bibnamefont
  {Watanabe}}, \bibinfo {author} {\bibfnamefont {T.}~\bibnamefont {Taniguchi}},
  \bibinfo {author} {\bibfnamefont {T.}~\bibnamefont {Ihn}}, \ and\ \bibinfo
  {author} {\bibfnamefont {K.}~\bibnamefont {Ensslin}},\ }\href@noop {}
  {\bibfield  {journal} {\bibinfo  {journal} {Applied Physics Letters}\
  }\textbf {\bibinfo {volume} {112}} (\bibinfo {year} {2018})}\BibitemShut
  {NoStop}%
\bibitem [{\citenamefont {Wang}\ \emph {et~al.}(2018)\citenamefont {Wang},
  \citenamefont {Greve}, \citenamefont {Jauregui}, \citenamefont {Sushko},
  \citenamefont {High}, \citenamefont {Zhou}, \citenamefont {Scuri},
  \citenamefont {Taniguchi}, \citenamefont {Watanabe}, \citenamefont {Lukin},
  \citenamefont {Park},\ and\ \citenamefont {Kim}}]{Wang2018}%
  \BibitemOpen
  \bibfield  {author} {\bibinfo {author} {\bibfnamefont {K.}~\bibnamefont
  {Wang}}, \bibinfo {author} {\bibfnamefont {K.~D.}\ \bibnamefont {Greve}},
  \bibinfo {author} {\bibfnamefont {L.~A.}\ \bibnamefont {Jauregui}}, \bibinfo
  {author} {\bibfnamefont {A.}~\bibnamefont {Sushko}}, \bibinfo {author}
  {\bibfnamefont {A.}~\bibnamefont {High}}, \bibinfo {author} {\bibfnamefont
  {Y.}~\bibnamefont {Zhou}}, \bibinfo {author} {\bibfnamefont {G.}~\bibnamefont
  {Scuri}}, \bibinfo {author} {\bibfnamefont {T.}~\bibnamefont {Taniguchi}},
  \bibinfo {author} {\bibfnamefont {K.}~\bibnamefont {Watanabe}}, \bibinfo
  {author} {\bibfnamefont {M.~D.}\ \bibnamefont {Lukin}}, \bibinfo {author}
  {\bibfnamefont {H.}~\bibnamefont {Park}}, \ and\ \bibinfo {author}
  {\bibfnamefont {P.}~\bibnamefont {Kim}},\ }\href@noop {} {\bibfield
  {journal} {\bibinfo  {journal} {Nature Nanotechnology}\ }\textbf {\bibinfo
  {volume} {13}} (\bibinfo {year} {2018})}\BibitemShut {NoStop}%
\bibitem [{\citenamefont {Davari}\ \emph {et~al.}(2020)\citenamefont {Davari},
  \citenamefont {Stacy}, \citenamefont {Mercado}, \citenamefont {Tull},
  \citenamefont {Basnet}, \citenamefont {Pandey}, \citenamefont {Watanabe},
  \citenamefont {Taniguchi}, \citenamefont {Hu},\ and\ \citenamefont
  {Churchill}}]{Davari2020}%
  \BibitemOpen
  \bibfield  {author} {\bibinfo {author} {\bibfnamefont {S.}~\bibnamefont
  {Davari}}, \bibinfo {author} {\bibfnamefont {J.}~\bibnamefont {Stacy}},
  \bibinfo {author} {\bibfnamefont {A.}~\bibnamefont {Mercado}}, \bibinfo
  {author} {\bibfnamefont {J.}~\bibnamefont {Tull}}, \bibinfo {author}
  {\bibfnamefont {R.}~\bibnamefont {Basnet}}, \bibinfo {author} {\bibfnamefont
  {K.}~\bibnamefont {Pandey}}, \bibinfo {author} {\bibfnamefont
  {K.}~\bibnamefont {Watanabe}}, \bibinfo {author} {\bibfnamefont
  {T.}~\bibnamefont {Taniguchi}}, \bibinfo {author} {\bibfnamefont
  {J.}~\bibnamefont {Hu}}, \ and\ \bibinfo {author} {\bibfnamefont
  {H.}~\bibnamefont {Churchill}},\ }\href@noop {} {\bibfield  {journal}
  {\bibinfo  {journal} {Physical Review Applied}\ }\textbf {\bibinfo {volume}
  {13}} (\bibinfo {year} {2020})}\BibitemShut {NoStop}%
\bibitem [{\citenamefont {Rhodes}\ \emph {et~al.}(2019)\citenamefont {Rhodes},
  \citenamefont {Chae}, \citenamefont {Ribeiro-Palau},\ and\ \citenamefont
  {Hone}}]{rhodes2019disorder}%
  \BibitemOpen
  \bibfield  {author} {\bibinfo {author} {\bibfnamefont {D.}~\bibnamefont
  {Rhodes}}, \bibinfo {author} {\bibfnamefont {S.~H.}\ \bibnamefont {Chae}},
  \bibinfo {author} {\bibfnamefont {R.}~\bibnamefont {Ribeiro-Palau}}, \ and\
  \bibinfo {author} {\bibfnamefont {J.}~\bibnamefont {Hone}},\ }\href@noop {}
  {\bibfield  {journal} {\bibinfo  {journal} {Nature materials}\ }\textbf
  {\bibinfo {volume} {18}},\ \bibinfo {pages} {541} (\bibinfo {year}
  {2019})}\BibitemShut {NoStop}%
\bibitem [{\citenamefont {Plumadore}\ \emph {et~al.}(2020)\citenamefont
  {Plumadore}, \citenamefont {Baskurt}, \citenamefont {Boddison-Chouinard},
  \citenamefont {Lopinski}, \citenamefont {Modarresi}, \citenamefont {Potasz},
  \citenamefont {Hawrylak}, \citenamefont {Sahin}, \citenamefont {Peeters},\
  and\ \citenamefont {Luican-Mayer}}]{plumadore2020defects}%
  \BibitemOpen
  \bibfield  {author} {\bibinfo {author} {\bibfnamefont {R.}~\bibnamefont
  {Plumadore}}, \bibinfo {author} {\bibfnamefont {M.}~\bibnamefont {Baskurt}},
  \bibinfo {author} {\bibfnamefont {J.}~\bibnamefont {Boddison-Chouinard}},
  \bibinfo {author} {\bibfnamefont {G.}~\bibnamefont {Lopinski}}, \bibinfo
  {author} {\bibfnamefont {M.}~\bibnamefont {Modarresi}}, \bibinfo {author}
  {\bibfnamefont {P.}~\bibnamefont {Potasz}}, \bibinfo {author} {\bibfnamefont
  {P.}~\bibnamefont {Hawrylak}}, \bibinfo {author} {\bibfnamefont
  {H.}~\bibnamefont {Sahin}}, \bibinfo {author} {\bibfnamefont {F.~M.}\
  \bibnamefont {Peeters}}, \ and\ \bibinfo {author} {\bibfnamefont
  {A.}~\bibnamefont {Luican-Mayer}},\ }\href {\doibase
  10.1103/PhysRevB.102.205408} {\bibfield  {journal} {\bibinfo  {journal}
  {Phys. Rev. B}\ }\textbf {\bibinfo {volume} {102}},\ \bibinfo {pages}
  {205408} (\bibinfo {year} {2020})}\BibitemShut {NoStop}%
\bibitem [{\citenamefont {Song}\ \emph
  {et~al.}(2015{\natexlab{a}})\citenamefont {Song}, \citenamefont {Liu},
  \citenamefont {Mosallanejad}, \citenamefont {You}, \citenamefont {Han},
  \citenamefont {Chen}, \citenamefont {Li}, \citenamefont {Cao}, \citenamefont
  {Xiao}, \citenamefont {Guo},\ and\ \citenamefont {Guo}}]{Guo1.2015}%
  \BibitemOpen
  \bibfield  {author} {\bibinfo {author} {\bibfnamefont {X.-X.}\ \bibnamefont
  {Song}}, \bibinfo {author} {\bibfnamefont {D.}~\bibnamefont {Liu}}, \bibinfo
  {author} {\bibfnamefont {V.}~\bibnamefont {Mosallanejad}}, \bibinfo {author}
  {\bibfnamefont {J.}~\bibnamefont {You}}, \bibinfo {author} {\bibfnamefont
  {T.-Y.}\ \bibnamefont {Han}}, \bibinfo {author} {\bibfnamefont {D.-T.}\
  \bibnamefont {Chen}}, \bibinfo {author} {\bibfnamefont {H.-O.}\ \bibnamefont
  {Li}}, \bibinfo {author} {\bibfnamefont {G.}~\bibnamefont {Cao}}, \bibinfo
  {author} {\bibfnamefont {M.}~\bibnamefont {Xiao}}, \bibinfo {author}
  {\bibfnamefont {G.-C.}\ \bibnamefont {Guo}}, \ and\ \bibinfo {author}
  {\bibfnamefont {G.-P.}\ \bibnamefont {Guo}},\ }\href@noop {} {\bibfield
  {journal} {\bibinfo  {journal} {Nanoscale}\ }\textbf {\bibinfo {volume}
  {7}},\ \bibinfo {pages} {16867} (\bibinfo {year}
  {2015}{\natexlab{a}})}\BibitemShut {NoStop}%
\bibitem [{\citenamefont {Song}\ \emph
  {et~al.}(2015{\natexlab{b}})\citenamefont {Song}, \citenamefont {Zhang},
  \citenamefont {You}, \citenamefont {Liu}, \citenamefont {Li}, \citenamefont
  {Cao}, \citenamefont {Xiao},\ and\ \citenamefont {Guo}}]{Guo2.2015}%
  \BibitemOpen
  \bibfield  {author} {\bibinfo {author} {\bibfnamefont {X.-X.}\ \bibnamefont
  {Song}}, \bibinfo {author} {\bibfnamefont {Z.-Z.}\ \bibnamefont {Zhang}},
  \bibinfo {author} {\bibfnamefont {J.}~\bibnamefont {You}}, \bibinfo {author}
  {\bibfnamefont {D.}~\bibnamefont {Liu}}, \bibinfo {author} {\bibfnamefont
  {H.-O.}\ \bibnamefont {Li}}, \bibinfo {author} {\bibfnamefont
  {G.}~\bibnamefont {Cao}}, \bibinfo {author} {\bibfnamefont {M.}~\bibnamefont
  {Xiao}}, \ and\ \bibinfo {author} {\bibfnamefont {G.-P.}\ \bibnamefont
  {Guo}},\ }\href@noop {} {\bibfield  {journal} {\bibinfo  {journal}
  {Scientific Reports}\ }\textbf {\bibinfo {volume} {5}} (\bibinfo {year}
  {2015}{\natexlab{b}})}\BibitemShut {NoStop}%
\bibitem [{\citenamefont {Wang}\ \emph {et~al.}(2013)\citenamefont {Wang},
  \citenamefont {Meric}, \citenamefont {Huang}, \citenamefont {Gao},
  \citenamefont {Gao}, \citenamefont {Tran}, \citenamefont {Taniguchi},
  \citenamefont {Watanabe}, \citenamefont {Campos}, \citenamefont {Muller},
  \citenamefont {Guo}, \citenamefont {Kim}, \citenamefont {Hone}, \citenamefont
  {Shepard},\ and\ \citenamefont {Dean}}]{Wang2013}%
  \BibitemOpen
  \bibfield  {author} {\bibinfo {author} {\bibfnamefont {L.}~\bibnamefont
  {Wang}}, \bibinfo {author} {\bibfnamefont {I.}~\bibnamefont {Meric}},
  \bibinfo {author} {\bibfnamefont {P.~Y.}\ \bibnamefont {Huang}}, \bibinfo
  {author} {\bibfnamefont {Q.}~\bibnamefont {Gao}}, \bibinfo {author}
  {\bibfnamefont {Y.}~\bibnamefont {Gao}}, \bibinfo {author} {\bibfnamefont
  {H.}~\bibnamefont {Tran}}, \bibinfo {author} {\bibfnamefont {T.}~\bibnamefont
  {Taniguchi}}, \bibinfo {author} {\bibfnamefont {K.}~\bibnamefont {Watanabe}},
  \bibinfo {author} {\bibfnamefont {L.~M.}\ \bibnamefont {Campos}}, \bibinfo
  {author} {\bibfnamefont {D.~A.}\ \bibnamefont {Muller}}, \bibinfo {author}
  {\bibfnamefont {J.}~\bibnamefont {Guo}}, \bibinfo {author} {\bibfnamefont
  {P.}~\bibnamefont {Kim}}, \bibinfo {author} {\bibfnamefont {J.}~\bibnamefont
  {Hone}}, \bibinfo {author} {\bibfnamefont {K.~L.}\ \bibnamefont {Shepard}}, \
  and\ \bibinfo {author} {\bibfnamefont {C.~R.}\ \bibnamefont {Dean}},\
  }\href@noop {} {\bibfield  {journal} {\bibinfo  {journal} {Science}\ }\textbf
  {\bibinfo {volume} {342}} (\bibinfo {year} {2013})}\BibitemShut {NoStop}%
\bibitem [{\citenamefont {Boddison-Chouinard}\ \emph
  {et~al.}(2019{\natexlab{a}})\citenamefont {Boddison-Chouinard}, \citenamefont
  {Plumadore},\ and\ \citenamefont {Luican-Mayer}}]{Boddison2019assembly}%
  \BibitemOpen
  \bibfield  {author} {\bibinfo {author} {\bibfnamefont {J.}~\bibnamefont
  {Boddison-Chouinard}}, \bibinfo {author} {\bibfnamefont {R.}~\bibnamefont
  {Plumadore}}, \ and\ \bibinfo {author} {\bibfnamefont {A.}~\bibnamefont
  {Luican-Mayer}},\ }\href@noop {} {\bibfield  {journal} {\bibinfo  {journal}
  {Journal of visualized experiments : JoVE}\ } (\bibinfo {year}
  {2019}{\natexlab{a}})}\BibitemShut {NoStop}%
\bibitem [{\citenamefont {Boddison-Chouinard}\ \emph
  {et~al.}(2019{\natexlab{b}})\citenamefont {Boddison-Chouinard}, \citenamefont
  {Scarfe}, \citenamefont {Watanabe}, \citenamefont {Taniguchi},\ and\
  \citenamefont {Luican-Mayer}}]{Boddison2019flat}%
  \BibitemOpen
  \bibfield  {author} {\bibinfo {author} {\bibfnamefont {J.}~\bibnamefont
  {Boddison-Chouinard}}, \bibinfo {author} {\bibfnamefont {S.}~\bibnamefont
  {Scarfe}}, \bibinfo {author} {\bibfnamefont {K.}~\bibnamefont {Watanabe}},
  \bibinfo {author} {\bibfnamefont {T.}~\bibnamefont {Taniguchi}}, \ and\
  \bibinfo {author} {\bibfnamefont {A.}~\bibnamefont {Luican-Mayer}},\
  }\href@noop {} {\bibfield  {journal} {\bibinfo  {journal} {Applied Physics
  Letters}\ }\textbf {\bibinfo {volume} {115}},\ \bibinfo {pages} {231603}
  (\bibinfo {year} {2019}{\natexlab{b}})}\BibitemShut {NoStop}%
\bibitem [{\citenamefont {Movva}\ \emph {et~al.}(2015)\citenamefont {Movva},
  \citenamefont {Rai}, \citenamefont {Kang}, \citenamefont {Kim}, \citenamefont
  {Fallahazad}, \citenamefont {Taniguchi}, \citenamefont {Watanabe},
  \citenamefont {Tutuc},\ and\ \citenamefont {Banerjee}}]{Movva2015}%
  \BibitemOpen
  \bibfield  {author} {\bibinfo {author} {\bibfnamefont {H.~C.~P.}\
  \bibnamefont {Movva}}, \bibinfo {author} {\bibfnamefont {A.}~\bibnamefont
  {Rai}}, \bibinfo {author} {\bibfnamefont {S.}~\bibnamefont {Kang}}, \bibinfo
  {author} {\bibfnamefont {K.}~\bibnamefont {Kim}}, \bibinfo {author}
  {\bibfnamefont {B.}~\bibnamefont {Fallahazad}}, \bibinfo {author}
  {\bibfnamefont {T.}~\bibnamefont {Taniguchi}}, \bibinfo {author}
  {\bibfnamefont {K.}~\bibnamefont {Watanabe}}, \bibinfo {author}
  {\bibfnamefont {E.}~\bibnamefont {Tutuc}}, \ and\ \bibinfo {author}
  {\bibfnamefont {S.~K.}\ \bibnamefont {Banerjee}},\ }\href@noop {} {\bibfield
  {journal} {\bibinfo  {journal} {ACS Nano}\ }\textbf {\bibinfo {volume} {9}}
  (\bibinfo {year} {2015})}\BibitemShut {NoStop}%
\bibitem [{\citenamefont {Goossens}\ \emph {et~al.}(2012)\citenamefont
  {Goossens}, \citenamefont {Calado}, \citenamefont {Barreiro}, \citenamefont
  {Watanabe}, \citenamefont {Taniguchi},\ and\ \citenamefont
  {Vandersypen}}]{Goossens2012}%
  \BibitemOpen
  \bibfield  {author} {\bibinfo {author} {\bibfnamefont {A.~M.}\ \bibnamefont
  {Goossens}}, \bibinfo {author} {\bibfnamefont {V.~E.}\ \bibnamefont
  {Calado}}, \bibinfo {author} {\bibfnamefont {A.}~\bibnamefont {Barreiro}},
  \bibinfo {author} {\bibfnamefont {K.}~\bibnamefont {Watanabe}}, \bibinfo
  {author} {\bibfnamefont {T.}~\bibnamefont {Taniguchi}}, \ and\ \bibinfo
  {author} {\bibfnamefont {L.~M.~K.}\ \bibnamefont {Vandersypen}},\ }\href@noop
  {} {\bibfield  {journal} {\bibinfo  {journal} {Applied Physics Letters}\
  }\textbf {\bibinfo {volume} {100}} (\bibinfo {year} {2012})}\BibitemShut
  {NoStop}%
\bibitem [{\citenamefont {Rosenberger}\ \emph {et~al.}(2018)\citenamefont
  {Rosenberger}, \citenamefont {Chuang}, \citenamefont {McCreary},
  \citenamefont {Hanbicki}, \citenamefont {Sivaram},\ and\ \citenamefont
  {Jonker}}]{Rosenberger2018}%
  \BibitemOpen
  \bibfield  {author} {\bibinfo {author} {\bibfnamefont {M.~R.}\ \bibnamefont
  {Rosenberger}}, \bibinfo {author} {\bibfnamefont {H.-J.}\ \bibnamefont
  {Chuang}}, \bibinfo {author} {\bibfnamefont {K.~M.}\ \bibnamefont
  {McCreary}}, \bibinfo {author} {\bibfnamefont {A.~T.}\ \bibnamefont
  {Hanbicki}}, \bibinfo {author} {\bibfnamefont {S.~V.}\ \bibnamefont
  {Sivaram}}, \ and\ \bibinfo {author} {\bibfnamefont {B.~T.}\ \bibnamefont
  {Jonker}},\ }\href@noop {} {\bibfield  {journal} {\bibinfo  {journal} {ACS
  Applied Materials \& Interfaces}\ }\textbf {\bibinfo {volume} {10}} (\bibinfo
  {year} {2018})}\BibitemShut {NoStop}%
\bibitem [{\citenamefont {Pradhan}\ \emph {et~al.}(2015)\citenamefont
  {Pradhan}, \citenamefont {Rhodes}, \citenamefont {Memaran}, \citenamefont
  {Poumirol}, \citenamefont {Smirnov}, \citenamefont {Talapatra}, \citenamefont
  {Feng}, \citenamefont {Perea-Lopez}, \citenamefont {Elias}, \citenamefont
  {Terrones} \emph {et~al.}}]{pradhan2015hall}%
  \BibitemOpen
  \bibfield  {author} {\bibinfo {author} {\bibfnamefont {N.}~\bibnamefont
  {Pradhan}}, \bibinfo {author} {\bibfnamefont {D.}~\bibnamefont {Rhodes}},
  \bibinfo {author} {\bibfnamefont {S.}~\bibnamefont {Memaran}}, \bibinfo
  {author} {\bibfnamefont {J.}~\bibnamefont {Poumirol}}, \bibinfo {author}
  {\bibfnamefont {D.}~\bibnamefont {Smirnov}}, \bibinfo {author} {\bibfnamefont
  {S.}~\bibnamefont {Talapatra}}, \bibinfo {author} {\bibfnamefont
  {S.}~\bibnamefont {Feng}}, \bibinfo {author} {\bibfnamefont {N.}~\bibnamefont
  {Perea-Lopez}}, \bibinfo {author} {\bibfnamefont {A.}~\bibnamefont {Elias}},
  \bibinfo {author} {\bibfnamefont {M.}~\bibnamefont {Terrones}},  \emph
  {et~al.},\ }\href@noop {} {\bibfield  {journal} {\bibinfo  {journal}
  {Scientific reports}\ }\textbf {\bibinfo {volume} {5}},\ \bibinfo {pages} {1}
  (\bibinfo {year} {2015})}\BibitemShut {NoStop}%
\bibitem [{\citenamefont {Young}\ \emph {et~al.}(2012)\citenamefont {Young},
  \citenamefont {Dean}, \citenamefont {Meric}, \citenamefont {Sorgenfrei},
  \citenamefont {Ren}, \citenamefont {Watanabe}, \citenamefont {Taniguchi},
  \citenamefont {Hone}, \citenamefont {Shepard},\ and\ \citenamefont
  {Kim}}]{Young2012}%
  \BibitemOpen
  \bibfield  {author} {\bibinfo {author} {\bibfnamefont {A.~F.}\ \bibnamefont
  {Young}}, \bibinfo {author} {\bibfnamefont {C.~R.}\ \bibnamefont {Dean}},
  \bibinfo {author} {\bibfnamefont {I.}~\bibnamefont {Meric}}, \bibinfo
  {author} {\bibfnamefont {S.}~\bibnamefont {Sorgenfrei}}, \bibinfo {author}
  {\bibfnamefont {H.}~\bibnamefont {Ren}}, \bibinfo {author} {\bibfnamefont
  {K.}~\bibnamefont {Watanabe}}, \bibinfo {author} {\bibfnamefont
  {T.}~\bibnamefont {Taniguchi}}, \bibinfo {author} {\bibfnamefont
  {J.}~\bibnamefont {Hone}}, \bibinfo {author} {\bibfnamefont {K.~L.}\
  \bibnamefont {Shepard}}, \ and\ \bibinfo {author} {\bibfnamefont
  {P.}~\bibnamefont {Kim}},\ }\href@noop {} {\bibfield  {journal} {\bibinfo
  {journal} {Physical Review B}\ }\textbf {\bibinfo {volume} {85}} (\bibinfo
  {year} {2012})}\BibitemShut {NoStop}%
\bibitem [{\citenamefont {Bieniek}\ \emph {et~al.}(2018)\citenamefont
  {Bieniek}, \citenamefont {Korkusi\ifmmode~\acute{n}\else \'{n}\fi{}ski},
  \citenamefont {Szulakowska}, \citenamefont {Potasz}, \citenamefont
  {Ozfidan},\ and\ \citenamefont {Hawrylak}}]{bieniek_hawrylak_prb2018}%
  \BibitemOpen
  \bibfield  {author} {\bibinfo {author} {\bibfnamefont {M.}~\bibnamefont
  {Bieniek}}, \bibinfo {author} {\bibfnamefont {M.}~\bibnamefont
  {Korkusi\ifmmode~\acute{n}\else \'{n}\fi{}ski}}, \bibinfo {author}
  {\bibfnamefont {L.}~\bibnamefont {Szulakowska}}, \bibinfo {author}
  {\bibfnamefont {P.}~\bibnamefont {Potasz}}, \bibinfo {author} {\bibfnamefont
  {I.}~\bibnamefont {Ozfidan}}, \ and\ \bibinfo {author} {\bibfnamefont
  {P.}~\bibnamefont {Hawrylak}},\ }\href {\doibase 10.1103/PhysRevB.97.085153}
  {\bibfield  {journal} {\bibinfo  {journal} {Phys. Rev. B}\ }\textbf {\bibinfo
  {volume} {97}},\ \bibinfo {pages} {085153} (\bibinfo {year}
  {2018})}\BibitemShut {NoStop}%
\bibitem [{\citenamefont {Alt{\i}nta{\c{s}}}\ \emph {et~al.}(2021)\citenamefont
  {Alt{\i}nta{\c{s}}}, \citenamefont {Bieniek}, \citenamefont {Dusko},
  \citenamefont {Korkusi{\'n}ski}, \citenamefont {Paw{\l}owski},\ and\
  \citenamefont {Hawrylak}}]{altintas_hawrylak_xxx2021}%
  \BibitemOpen
  \bibfield  {author} {\bibinfo {author} {\bibfnamefont {A.}~\bibnamefont
  {Alt{\i}nta{\c{s}}}}, \bibinfo {author} {\bibfnamefont {M.}~\bibnamefont
  {Bieniek}}, \bibinfo {author} {\bibfnamefont {A.}~\bibnamefont {Dusko}},
  \bibinfo {author} {\bibfnamefont {M.}~\bibnamefont {Korkusi{\'n}ski}},
  \bibinfo {author} {\bibfnamefont {J.}~\bibnamefont {Paw{\l}owski}}, \ and\
  \bibinfo {author} {\bibfnamefont {P.}~\bibnamefont {Hawrylak}},\ }\href@noop
  {} {\bibfield  {journal} {\bibinfo  {journal} {arXiv preprint
  arXiv:2106.15090}\ } (\bibinfo {year} {2021})}\BibitemShut {NoStop}%
\bibitem [{\citenamefont {Bieniek}\ \emph {et~al.}(2020)\citenamefont
  {Bieniek}, \citenamefont {Szulakowska},\ and\ \citenamefont
  {Hawrylak}}]{bieniek_hawrylak_prb2020}%
  \BibitemOpen
  \bibfield  {author} {\bibinfo {author} {\bibfnamefont {M.}~\bibnamefont
  {Bieniek}}, \bibinfo {author} {\bibfnamefont {L.}~\bibnamefont
  {Szulakowska}}, \ and\ \bibinfo {author} {\bibfnamefont {P.}~\bibnamefont
  {Hawrylak}},\ }\href {\doibase 10.1103/PhysRevB.101.035401} {\bibfield
  {journal} {\bibinfo  {journal} {Phys. Rev. B}\ }\textbf {\bibinfo {volume}
  {101}},\ \bibinfo {pages} {035401} (\bibinfo {year} {2020})}\BibitemShut
  {NoStop}%
\bibitem [{\citenamefont {Szulakowska}\ \emph {et~al.}(2020)\citenamefont
  {Szulakowska}, \citenamefont {Cygorek}, \citenamefont {Bieniek},\ and\
  \citenamefont {Hawrylak}}]{szulakowska_hawrylak_prb2020}%
  \BibitemOpen
  \bibfield  {author} {\bibinfo {author} {\bibfnamefont {L.}~\bibnamefont
  {Szulakowska}}, \bibinfo {author} {\bibfnamefont {M.}~\bibnamefont
  {Cygorek}}, \bibinfo {author} {\bibfnamefont {M.}~\bibnamefont {Bieniek}}, \
  and\ \bibinfo {author} {\bibfnamefont {P.}~\bibnamefont {Hawrylak}},\ }\href
  {\doibase 10.1103/PhysRevB.102.245410} {\bibfield  {journal} {\bibinfo
  {journal} {Phys. Rev. B}\ }\textbf {\bibinfo {volume} {102}},\ \bibinfo
  {pages} {245410} (\bibinfo {year} {2020})}\BibitemShut {NoStop}%
\bibitem [{\citenamefont {Raymond}\ \emph {et~al.}(2004)\citenamefont
  {Raymond}, \citenamefont {Studenikin}, \citenamefont {Sachrajda},
  \citenamefont {Wasilewski}, \citenamefont {Cheng}, \citenamefont {Sheng},
  \citenamefont {Hawrylak}, \citenamefont {Babinski}, \citenamefont {Potemski},
  \citenamefont {Ortner} \emph {et~al.}}]{raymond_hawrylak_prl2004}%
  \BibitemOpen
  \bibfield  {author} {\bibinfo {author} {\bibfnamefont {S.}~\bibnamefont
  {Raymond}}, \bibinfo {author} {\bibfnamefont {S.}~\bibnamefont {Studenikin}},
  \bibinfo {author} {\bibfnamefont {A.}~\bibnamefont {Sachrajda}}, \bibinfo
  {author} {\bibfnamefont {Z.}~\bibnamefont {Wasilewski}}, \bibinfo {author}
  {\bibfnamefont {S.-J.}\ \bibnamefont {Cheng}}, \bibinfo {author}
  {\bibfnamefont {W.}~\bibnamefont {Sheng}}, \bibinfo {author} {\bibfnamefont
  {P.}~\bibnamefont {Hawrylak}}, \bibinfo {author} {\bibfnamefont
  {A.}~\bibnamefont {Babinski}}, \bibinfo {author} {\bibfnamefont
  {M.}~\bibnamefont {Potemski}}, \bibinfo {author} {\bibfnamefont
  {G.}~\bibnamefont {Ortner}},  \emph {et~al.},\ }\href@noop {} {\bibfield
  {journal} {\bibinfo  {journal} {Physical review letters}\ }\textbf {\bibinfo
  {volume} {92}},\ \bibinfo {pages} {187402} (\bibinfo {year}
  {2004})}\BibitemShut {NoStop}%
\bibitem [{\citenamefont {Korkusinski}\ \emph {et~al.}(2003)\citenamefont
  {Korkusinski}, \citenamefont {Sheng},\ and\ \citenamefont
  {Hawrylak}}]{sheng_hawrylak_pss2003}%
  \BibitemOpen
  \bibfield  {author} {\bibinfo {author} {\bibfnamefont {M.}~\bibnamefont
  {Korkusinski}}, \bibinfo {author} {\bibfnamefont {W.}~\bibnamefont {Sheng}},
  \ and\ \bibinfo {author} {\bibfnamefont {P.}~\bibnamefont {Hawrylak}},\
  }\href@noop {} {\bibfield  {journal} {\bibinfo  {journal} {physica status
  solidi (b)}\ }\textbf {\bibinfo {volume} {238}},\ \bibinfo {pages} {246}
  (\bibinfo {year} {2003})}\BibitemShut {NoStop}%
\end{thebibliography}%

\end{document}